# Comment: Boosting Algorithms: Regularization, Prediction and Model Fitting

Andreas Buja, David Mease and Abraham J. Wyner

*Abstract.* The authors are doing the readers of *Statistical Science* a true service with a well-written and up-to-date overview of boosting that originated with the seminal algorithms of Freund and Schapire. Equally, we are grateful for high-level software that will permit a larger readership to experiment with, or simply apply, boosting-inspired model fitting. The authors show us a world of methodology that illustrates how a fundamental innovation can penetrate every nook and cranny of statistical thinking and practice. They introduce the reader to one particular interpretation of boosting and then give a display of its potential with extensions from classification (where it all started) to least squares, exponential family models, survival analysis, to base-learners other than trees such as smoothing splines, to degrees of freedom and regularization, and to fascinating recent work in model selection. The uninitiated reader will find that the authors did a nice job of presenting a certain coherent and useful interpretation of boosting. The other reader, though, who has watched the business of boosting for a while, may have quibbles with the authors over details of the historic record and, more importantly, over their optimism about the current state of theoretical knowledge. In fact, as much as "the statistical view" has proven fruitful, it has also resulted in some ideas about why boosting works that may be misconceived, and in some recommendations that may be misguided.

*Andreas Buja is Liem Sioe Liong/First Pacific Company Professor of Statistics, Statistics Department, The Wharton School, University of Pennsylvania, Philadelphia, Pennsylvania 19104-6340, USA e-mail: buja.at.wharton@gmail.com. David Mease is Assistant Professor, Department of Marketing and Decision Sciences, College of Business, San Jose State University, San Jose, California 95192-0069, USA e-mail: mease_d@cob.sjsu.edu. Abraham J. Wyner is Associate Professor, Statistics Department, The Wharton School, University of Pennsylvania, Philadelphia, Pennsylvania 19104-6340, USA e-mail: ajw@wharton.upenn.edu.*

## HISTORY OF "THE STATISTICAL VIEW" AND FIRST QUESTIONS

To get a sense of past history as well as of current ignorance, we must go back to the roots of boosting, which are in classification. On this way back, we will take the late Leo Breiman as our guide, because learning what he knew or did not know is instructive to this day.

Only a decade ago Freund and Schapire (1997, page 119), defined boosting as "converting a 'weak' PAC learning algorithm that performs just slightly better than random guessing into one with arbitrarily high accuracy." The assumptions underlying the quote imply that the classes are 100% separable and hence that classification solves basically a geometric problem. How else would one interpret "arbitrarily high accuracy" other than implying a zero Bayes error? See Breiman's (1998, Appendix) patient but firm comments on this point. To a statistician the early literature on boosting was an interesting mix of creativity, technical bravado, and statistically unrealistic assumptions inspired by the PAC learning framework. Yet, in as far as machine learners relied on Vapnik's random sampling assumption and his allowance for overlapping classes, they had in hand the seeds for a fundamentally statistical treatment of boosting, at least in theory.

By now, statistical views of boosting have existed for a number of years, and they are mostly due to statisticians. One such view is due to Friedman, Hastie and Tibshirani (2000) who propose that boosting is stagewise additive model fitting. Equivalent to stagewise additive fitting is Bühlmann and Hothorn's notion of fitting by gradient descent in function space, theirs being a more mathematical than statistical terminology. Bühlmann and Hothorn attribute the view of boosting as functional gradient descent (FGD) to Breiman, but in this they are factually inaccurate. Of the two articles they cite, "Arcing Classifiers" (Breiman, 1998) has nothing to do with optimization. Here is Breiman's famous praise of boosting algorithms as "the most accurate ... off-the-shelf classifiers on a wide variety of data sets." The article is important, but not as an ancestor of the "statistical view" of boosting as we will see below. A better candidate is Bühlmann and Hothorn's other reference, "Prediction Games and Arcing Algorithms" (Breiman, 1999). A closer reading shows, however, that it is an ancestor, not a founder, of a statistical view of boosting, even though here is the first interpretation of AdaBoost as minimization of an exponential criterion. Borrowing from Freund and Schapire (1996), Breiman's approach is not statistical but game-theoretic, hence he justifies fitting base learners not with gradient descent but with the minimax theorem. He stylizes the problem to selecting among finitely many fixed base learners, thereby removing the functional aspect. His calculations are on training samples, not populations, and hence they never reveal what is being estimated. In his pre-2000 work one will find neither the terms "functional" and "gradient" nor a concept of boosting as model fitting and estimation. These facts stand against Mason et al.'s (2000, Section 2.1) attribution of "gradient descent in function space" to Breiman, against Breiman (2000a, 2004) himself when he links FGD to Breiman (1999, 1997), and now against Bühlmann and Hothorn.

For a statistical view of boosting, the dam really broke in 1998 with a report by Friedman, Hastie and Tibshirani (2000, based on a 1998 report; "FHT (2000)" henceforth). Around that time, others had also picked up on the exponential criterion and its minimization, including Mason et al. (2000) and Schapire and Singer (1999), but it was FHT (2000) whose simple population calculations established the meaning of boosting as model fitting in the following sense: Boosting creates linear combinations of base learners (called "weighted votes" in machine learning) that are estimates of half the logit of the underlying conditional class probabilities, $P(Y=1|x)$. In this view, boosting could suddenly be seen as class probability estimation in the conditional Bernoulli model, and consequently FHT's (2000) first order of business was to create LogitBoost by replacing exponential loss with the loss function that is natural to statisticians, the negative log-likelihood of the Bernoulli model (= "log-loss"). FHT (2000) also replaced boosting's reweighting with the reweighting that statisticians have known for decades, iteratively reweighted least squares, to implement Newton descent/Fisher scoring. In this clean picture, AdaBoost estimates half the logit, LogitBoost estimates the logit, both by stagewise fitting, but by different approaches to the functional gradient that produces the additive terms. Going yet further, Friedman (2001, based on a 1999 report) discarded weighting altogether by approximating gradients with plain least squares. These innovations had been absorbed as early as 1999 by the newly minted Ph.D. Greg Ridgeway (1999) who presented an excellent piece on "The State of Boosting" that included a survey of these yet-to-be-published developments as well as his own work on boosting for exponential family and survival regression. Thus the new view of boosting as model fitting developed in a short period between the middle of 1998 and early 1999 and bore fruit instantly before any of it had appeared in print.

It is Friedman's (2001) gradient boosting that Bühlmann and Hothorn now call "the generic FGD

or boosting algorithm" (Section 2.1). This promotion of one particular algorithm to a standard could give rise to misgivings among the originators of boosting because the original discrete AdaBoost (Section 1.2) is not even a special case of gradient boosting. There exists, however, a version of gradient descent that contains AdaBoost as a special case: it is alluded to in Section 2.1.1 and appears in Mason et al. (2000, Section 3), FHT (2000, Section 4.1) and Breiman (2000a; 2004, Sections 2.2, 4.1). Starting with the identity

$$\frac{\partial}{\partial t}\bigg|_{t=0} \sum_i \rho(Y_i, f(X_i) + tg(X_i))$$
$$= \sum_i \rho'(Y_i, f(X_i)) g(X_i)$$

($\rho'$ = the partial w.r.t. the second argument), find steepest descent directions by minimizing the right-hand expression with regard to $g(X)$. Minimization in this case is not generally well defined, because it typically produces $-\infty$ unless the permissible directions $g(X)$ are bounded (Ridgeway, 2000). One way to bound $g(X)$ is by confining it to classifiers ($g(X) \in \{-1, +1\}$), in which case gradient descent on the exponential loss function $\rho = \exp(-Y_i f(X_i))$ ($Y_i = \pm 1$) yields discrete AdaBoost. Instead of bounding of $g(X)$, Ridgeway (2000) pointed out that the above ill-posed gradient minimization could be regularized by adding a quadratic penalty $Q(\mathbf{g}) = \sum_i g(X_i)^2/2$ to the right-hand side, only to arrive at a criterion that, after quadratic completion, produces Friedman's (2001) least squares gradient boosting:

$$\sum_i ((-\rho'(Y_i, f(X_i))) - g(X_i))^2.$$

We may wonder what, other than algebraic convenience, makes $\sum_i g(X_i)^2/2$ the penalty of choice. A mild modification is $Q(\mathbf{g}) = 1/(2c) \sum_i g(X_i)^2$ with $c > 0$ as a penalty parameter; quadratic completion results in the least squares criterion

$$\sum_i ((-c\rho'(Y_i, f(X_i))) - g(X_i))^2,$$

which shows that for small $c$ its minimization yields Friedman's step size shrinkage. The choice

$$Q(\mathbf{g}) = \sum_i \rho''(Y_i, f(X_i)) g(X_i)^2/2$$

has the particular justification that it provides a second-order approximation to the loss function, and hence its minimization generates Newton descent/Fisher scoring as used in FHT's LogitBoost. For comparison, gradient descent uses $-\rho'(Y_i, f(X_i))$ as the working response in an unweighted least squares problem, whereas Newton descent uses $(-\rho'/\rho'')(Y_i, f(X_i))$ as the working response in a weighted least squares problem with weights $\rho''(Y_i, f(X_i))$. In view of these choices, we may ask Bühlmann and Hothorn whether there are deeper reasons for their advocacy of Friedman's gradient descent as the boosting standard. Friedman's intended applications included $L_1$- and Huber M-estimation, in which case second derivatives are not available. In many other cases, though, including exponential and logistic loss and the likelihood of any exponential family model, second derivatives are available, and we should expect some reasoning from Bühlmann and Hothorn for abandoning entrenched statistical practice.

## LIMITATIONS OF "THE STATISTICAL VIEW" OF BOOSTING

While the statistical view of boosting as model fitting is truly a breakthrough and has proven extremely fruitful in spawning new boosting methodologies, one should not ignore that it has also caused misconceptions, in particular in classification. For example, the idea that boosting implicitly estimates conditional class probabilities turns out to be wrong in practice. Both AdaBoost and LogitBoost are primarily used for classification, not class probability estimation, and in so far as they produce successful classifiers in practice, they also produce extremely overfitted estimates of conditional class probabilities, namely, values near zero and 1. In other words, it would be a mistake to assume that in order to successfully classify, one should look for accurate class probability estimates. Successful classification cannot be reduced to successful class probability estimation, and some published theoretical work is flawed because of doing just that. Bühlmann and Hothorn allude to these problems in Section 1.3, but they do not discuss them. It would be helpful if they summarized for us the state of statistical theory in explaining successful classification without committing the fallacy of reducing it to successful class probability estimation.

There have been some misunderstandings in the literature about an alleged superiority of LogitBoost over AdaBoost for class probability estimation. No such thing can be asserted to date. Both produce scores that are in theory estimates of $P(Y = 1|x)$ when passed through an inverse link function. Both could be used for class probability estimation if properly regularized—at the cost of deteriorating classification performance. Bühlmann and Hothorn's list of reasons for preferring log-loss over exponential loss (Section 3.2.1) might cater to some of the more common misconceptions: log-loss "(i) ...yields probability estimates"—so does exponential loss; both do so in theory but not in practice, unless either loss function is suitably regularized; "(ii) it is a monotone loss function of the margin"—so is exponential loss; "(iii) it grows linearly as the margin... tends to $-\infty$, unlike the exponential loss"—true, but when they add "The third point reflects a robustness aspect: it is similar to Huber's loss function," they are overstepping the boundaries of today's knowledge. Do we know that there even exists a robustness issue? Unlike quantitative responses, binary responses have no problem of vertically outlying values. The stronger growth of the exponential loss only implies greater penalties for strongly misclassified cases, and why should this be detrimental? It appears that there is currently no theory that allows us to recommend log-loss over exponential loss or vice versa, or to choose from the larger class of proper scoring rules described by Buja et al. (2005). If Bühlmann and Hothorn have a stronger argument to make, it would be most welcome.

For our next point, we return to Breiman's (1998) article because its main message is a heresy in light of today's "statistical view" of boosting. He writes: "The main effect of both bagging and arcing is to reduce variance" (page 802; "arcing" = Breiman's term for boosting). This was written before his discovery of boosting's connection with exponential loss, from a performance-oriented point of view informed by a bias-variance decomposition he devised for classification. It was also before the advent of the "statistical view" and its "low-variance principle," which explains Breiman's use of the full CART algorithm as the base learner, following earlier examples in machine learning that used the full C4.5 algorithm.

Then Breiman (1999, page 1494) dramatically reverses himself in response to learning that "Schapire et al. (1997) [(1998)] gave examples of data where two-node trees (stumps) had high bias and the main effect of AdaBoost was to reduce the bias." This work of Breiman's makes fascinating reading because of its perplexed tone and its admission in the Conclusions section (page 1506) that "the results leave us in a quandary," and "the laboratory results for various arcing algorithms are excellent, but the theory is in disarray." His important discovery that AdaBoost can be interpreted as the minimizer of an exponential criterion happens on the side line of an argument with Schapire and Freund about the deficiencies of VC- and margin-based arguments for explaining boosting. Yet, thereafter Breiman no longer cites his 1998 *Annals* article in a substantive way, and he, too, submits to the idea that the complexity of base learners needs to be controlled. Today we seem to be sworn in on base learners that are weak in the sense of having low complexity, high bias (for most data) and low variance, and accordingly Bühlmann and Hothorn exhort us to adopt the "low-variance principle" (Section 4.4). What PAC theory used to call "weak learner" is now statistically re-interpreted as "low-variance learner." In this we miss out on the other possible cause of weakness, which is high variance. As much as underfitting calls for bias reduction, overfitting calls for variance reduction. Some varieties of boosting may be able to achieve both, whereas current theories and the "statistical view" in general obsess with bias. Against today's consensus we need to draw attention again to the earlier Breiman (1998) to remind us of his and others' favorable experiences with boosting of high-variance base learners such as CART and C4.5. It was in the high-variance case that Breiman issued his praise of boosting, and it is this case that seems to be lacking theoretical explanation. Obviously, high-variance base learners cannot be analyzed with a heuristic such as in Bühlmann and Hothorn's Section 5.1 (from Bühlmann and Yu, 2003) for $L_2$ boosting which only transfers variability from residuals to fits and never the other way round. Ideally, we would have a single approach that automatically reduces bias when necessary and variance when necessary. That such could be the case for some versions of AdaBoost was still in the back of Breiman's mind, and it is now explicitly asserted by Amit and Blanchard (2001), not only for AdaBoost but for a large class of ensemble methods. Is this a statistical jackpot, and we are not realizing it because we are missing the theory to comprehend it?

After his acquiescence to low-complexity base learners and regularization, Breiman still uttered occasionally a discordant view, as in his work on random forests (Breiman, 1999b, page 3) where he conjectured: "Adaboost has no random elements ... But just as a deterministic random number generator can give a good imitation of randomness, my belief is that in its later stages Adaboost is emulating a random forest." If his intuition is on target, then we may want to focus on randomized versions of boosting for variance reduction, both in theory and practice. On the practical side, Friedman (2002, based on a report of 1999) took a leaf out of Breiman's book and found that restricting boosting iterations to random subsamples improved performance in the vast majority of scenarios he examined. The abstract of Friedman's article ends on this note: "This randomized approach also increases robustness against overcapacity of the base learner," that is, against overfitting by a high-variance base learner. This simple yet powerful extension of functional gradient descent is not mentioned by Bühlmann and Hothorn. Yet, Breiman's and Friedman's work seems to point to a statistical jackpot outside the "statistical view."

## LIMITATIONS OF "THE STATISTICAL VIEW" OF BOOSTING EXEMPLIFIED

In the previous section we outlined limitations of the prevalent "statistical view" of boosting by following some of boosting's history and pointing to misconceptions and blind spots in "the statistical view." In this section we will sharpen our concerns based on an article, "Evidence Contrary to the Statistical View of Boosting," by two of us (Mease and Wyner, 2007, "MW (2007)" henceforth), to appear in the *Journal of Machine Learning Research* (JMLR). Understandably this article was not known to Bühlmann and Hothorn at the time when they wrote theirs, as we were not aware of theirs when we wrote ours. Since these two works represent two contemporary contesting views, we feel it is of interest to discuss the relationship further. Specifically, in this section we will draw connections between statements made in Bühlmann and Hothorn's article and evidence against these statements presented in our JMLR article. In what follows, we provide a list of five beliefs central to the statistical view of boosting. For each of these, we cite specific statements in the Bühlmann–Hothorn article that reflect these beliefs. Then we briefly discuss empirical evidence presented in our JMLR article that calls these beliefs into question. The discussion is now limited to two-class classification where boosting's peculiarities are most in focus. The algorithm we use is "discrete AdaBoost."

### Statistical Perspective on Boosting Belief #1: Stumps Should Be Used for Additive Bayes Decision Rules

In their Section 4.3 Bühlmann and Hothorn reproduce the following argument from FHT (2000): "When using stumps ... the boosting estimate will be an additive model in the original predictor variables, because every stump-estimate is a function of a single predictor variable only. Similarly, boosting trees with (at most) $d$ terminal nodes results in a nonparametric model having at most interactions of order $d - 2$. Therefore, if we want to constrain the degree of interactions, we can easily do this by constraining the (maximal) number of nodes in the base procedure." In Section 4.4 they suggest to "choose the base procedure (having the desired structure) with low variance at the price of larger estimation bias." As a consequence, if one decides that the desired structure is an additive model, the best choice for a base learner would be stumps. While this belief certainly is well accepted in the statistical community, practice suggests otherwise. It can easily be shown through simulation that boosted stumps often perform substantially worse than larger trees even when the true classification boundaries can be described by an additive function. A striking example is given in Section 3.1 of our JMLR article. In this simulation not only do stumps give a higher misclassification error (even with the optimal stopping time), they also exhibit substantial overfitting while the larger trees show no signs of overfitting in the first 1000 iterations and lead to a much smaller hold-out misclassification error.

### Statistical Perspective on Boosting Belief #2: Early Stopping Should Be Used to Prevent Overfitting

In Section 1.3 Bühlmann and Hothorn tell us that "it is clear nowadays that AdaBoost and also other boosting algorithms are overfitting eventually, and early stopping is necessary." This statement is extremely broad and contradicts Breiman (2000b) who wrote, based on empirical evidence, that "A crucial property of AdaBoost is that it almost never overfits the data no matter how many iterations it is run." The contrast might suggest that in the seven years since, there has been theory or further empirical evidence to verify that overfitting will happen eventually in all of the instances on which Breiman based his claim. No such theory exists and empirical examples of overfitting are rare, especially for relatively high-variance base learners. Ironically, stumps with low variance seem to be more prone to overfitting than base learners with high variance. Also, some examples of overfitting in the literature are quite artificial and often employ algorithms that bear little resemblance to the original AdaBoost algorithm. On the other hand, examples for which overfitting is not observed are abundant, and a number of such examples are given in our JMLR article. If overfitting is judged with respect to misclassification error, not only does the empirical evidence suggest early stopping is not necessary in most applications of AdaBoost, but early stopping can degrade performance. Another matter is overfitting in terms of the conditional class probabilities as measured by the surrogate loss function (exponential loss, negative log-likelihood, proper scoring rules in general; see Buja et al., 2005). Class probabilities tend to overfit rapidly and drastically, while hold-out misclassification errors keep improving.

### Statistical Perspective on Boosting Belief #3: Shrinkage Should Be Used to Prevent Overfitting

Shrinkage in boosting is the practice of using a step-length factor smaller than 1. It is discussed in Section 2.1 where the authors write the following: "The choice of the step-length factor $\nu$ in step 4 is of minor importance, as long as it is 'small' such as $\nu = 0.1$. A smaller value of $\nu$ typically requires a larger number of boosting iterations and thus more computing time, while the predictive accuracy has been empirically found to be potentially better and almost never worse when choosing $\nu$ 'sufficiently small' (e.g., $\nu = 0.1$)." With regard to AdaBoost, these statements are generally not true. In fact, not only does shrinkage often not improve performance, it can lead to overfitting in cases in which AdaBoost otherwise would not overfit. An example can be found in Section 3.7 of our JMLR article.

### Statistical Perspective on Boosting Belief #4: Boosting is Estimating Probabilities

In Section 3.1 Bühlmann and Hothorn present the usual probability estimates for AdaBoost that emerge from the "statistical view," mentioning that "the reason for constructing these probability estimates is based on the fact that boosting with a suitable stopping iteration is consistent." While the "statistical view" of boosting does in fact suggest this mapping produces estimates of the class probabilities, they tend to produce uncompetitive classification if stopped early, or else vastly overfitted class probabilities if stopped late. We do caution against their use in the article cited by the authors (Mease, Wyner, Buja, 2007). In that article we further show that simple approaches based on over- and undersampling yield class probability estimates that perform quite well. In MW (2007) we give a simple example for which the true conditional probabilities of class 1 are either 0.1 or 0.9, yet the probability estimates quickly diverge to values smaller than 0.01 and larger than 0.99 well before the classification rule has approached its optimum. This behavior is typical.

### Statistical Perspective on Boosting Belief #5: Regularization Should Be Based on the Loss Function

In Section 5.4 the authors suggest one can "use information criteria for estimating a good stopping iteration." One of these criteria suggested for the classification problem is an AIC- or BIC-penalized negative binomial log-likelihood. A problem with Bühlmann and Hothorn's presentation is that they do not explain whether their recommendation is intended for estimating conditional class probabilities or for classification. In the case of classification, readers should be warned that the recommendation will produce inferior performance for reasons explained earlier: Boosting iterations keep improving in terms of hold-out misclassification error while class probabilities are being overfitted beyond reason. While early stopping based on penalized likelihoods might produce reasonable values for conditional class probabilities, the resulting classifiers would be entirely uncompetitive in terms of hold-out misclassification error. In our two JMLR articles (Mease et al., 2007; MW, 2007) we provide a number of examples in which the hold-out misclassification error decreases throughout while the hold-out binomial log-likelihood and similar measures deteriorate throughout. This would suggest that the "good stopping iteration" is the very first iteration, when in fact for classification the best iteration is the last iteration which is at least 800 in all examples.

## WHAT IS THE ROLE OF THE SURROGATE LOSS FUNCTION?

In this last section we wish to further muddy our view of the role of surrogate loss functions as well as the issues of step-size selection and early stopping.

COMMENT 7

Drawing on Wyner (2003), we consider a modification of AdaBoost that doubles the step size relative to the standard AdaBoost algorithm:

$$\alpha^{[m]} = 2\log\left(\frac{1-err^{[m]}}{err^{[m]}}\right).$$

The additional factor of 2 of course does not simply double all the coefficients because it affects the reweighting at each iteration: starting with the second iteration, raw and modified AdaBoost will use different sets of weights, hence the fitted base learners will differ.

As can be seen from the description of the AdaBoost algorithm in Bühlmann and Hothorn's Section 1.2, doubling the step size amounts to using the square of the weight multiplier in each iteration. It is obvious that the modified AdaBoost uses a more aggressive reweighting strategy because, relatively speaking, squaring makes small weights smaller and large weights larger. Just the same, modified AdaBoost is a reweighting algorithm that is very similar to the original AdaBoost, and it is not a priori clear which of the two algorithms is going to be the more successful one.

It is obvious, however, that modified AdaBoost does strange things in terms of the exponential loss. We know that the original AdaBoost's step-size choice is the minimizer in a line search of the exponential loss in the direction of the fitted base learner. Doubling the step size overshoots the line search by not descending to the valley but re-ascending on the opposite slope of the exponential loss function. Even more is known: Wyner (2003) showed that the modified algorithm re-ascends in such a way that the exponential loss is the same as in the previous iteration! In other words, the value of the exponential loss remains constant across iterations. Still more is known: it can be shown that there does not exist any loss function for which modified AdaBoost yields the minimizer of a line search.

Are we to conclude that modified AdaBoost must perform badly? This could not be further from the truth: with C4.5 as the base learner, misclassification errors tend to approach zero quickly on the training data and tend to decrease long thereafter on the hold-out data, just as in AdaBoost. As to the bottom line, the modified algorithm is comparable to AdaBoost: hold-out misclassification errors after over 200 iterations are not identical but similar on average to AdaBoost's (Wyner, 2003, Figures 1–3). What is the final analysis of these facts? At a minimum, we can say that they throw a monkey wrench into the tidy machinery of the "statistical view of boosting."

## CONCLUSIONS

There is something missing in the "statistical view of boosting," and what is missing results in misguided recommendations. By guiding us toward high-bias/low-variance/low-complexity base learners for boosting, the "view" misses out on the power of boosting low-bias/high-variance/high-complexity base learners such as C4.5 and CART. It was in this context that boosting had received its original praise in the statistics world (Breiman, 1998). The situation in which the "statistical view" finds itself is akin to the joke in which a man looks for the lost key under the street light even though he lost it in the dark. The "statistical view" uses the ample light of traditional model fitting that is based on predictors with weak explanatory power. A contrasting view, pioneered by the earlier Breiman as well as Amit and Geman (1997) and associated with the terms "bagging" and "random forests," assumes predictor sets so rich that they overfit and require variance- instead of bias-reduction. Breiman's (1998) early view was that boosting is like bagging, only better, in its ability to reduce variance. By not accounting for variance reduction, the "statistical view" guides us into a familiar corner where there is plenty of light but where we might be missing out on more powerful fitting technology.

## REFERENCES

AMIT, Y. and BLANCHARD, G. (2001). Multiple randomized classifiers: MRCL. Technical report, Univ. Chicago.

AMIT, Y. and GEMAN, D. (1997). Shape quantization and recognition with randomized trees. *Neural Computation* **9** 1545–1588.

BREIMAN, L. (1997). Arcing the edge. Technical Report 486, Dept. Statistics, Univ. California. Available at www.stat.berkeley.edu.

BREIMAN, L. (1998). Arcing classifiers (with discussion). *Ann. Statist.* **26** 801–849. MR1635406

BREIMAN, L. (1999). Random forests—Random features. Technical Report 567, Dept. Statistics, Univ. California. Available at www.stat.berkeley.edu.

BREIMAN, L. (2000a). Some infinity theory for predictor ensembles. Technical Report 577, Dept. Statistics, Univ. California. Available at www.stat.berkeley.edu.

BREIMAN, L. (2000b). Discussion of "Additive logistic regression: A statistical view of boosting," by J. Friedman, T. Hastie and R. Tibshirani. *Ann. Statist.* **28** 374–377. MR1790002

Breiman, L. (2004). Population theory for boosting ensembles. *Ann. Statist.* **32** 1–11. MR2050998

Buja, A., Stuetzle, W. and Shen, Y. (2005). Loss functions for binary class probability estimation: Structure and applications. Technical report, Univ. Washington. Available at http://www.stat.washington.edu/wxs/Learning-papers/paper-proper-scoring.pdf.

Bühlmann, P. and Yu, B. (2003). Boosting with the $L_2$ loss: Regression and classification. *J. Amer. Statist. Assoc.* **98** 324–339. MR1995709

Friedman, J. (2001). Greedy function approximation: A gradient boosting machine. *Ann. Statist.* **29** 1189–1232. MR1873328

Friedman, J. H. (2002). Stochastic gradient boosting. *Comput. Statist. Data Anal.* **38** 367–378. MR1884869

Friedman, J. H., Hastie, T. and Tibshirani, R. (2000). Additive logistic regression: A statistical view of boosting (with discussion). *Ann. Statist.* **38** 367–378. MR1790002

Freund, Y. and Schapire, R. (1996). Experiments with a new boosting algorithm. In *Proceedings of the Thirteenth International Conference on Machine Learning*. Morgan Kaufmann, San Francisco, CA.

Freund, Y. and Schapire, R. (1997). A decision-theoretic generalization of on-line learning and an application to boosting. *J. Comput. System Sci.* **55** 119–139. MR1473055

Mason, L., Baxter, J., Bartlett, P. and Frean, M. (2000). Functional gradient techniques for combining hypotheses. In *Advances in Large Margin Classifiers* (A. Smola, P. Bartlett, B. Schölkopf and D. Schuurmans, eds.). MIT Press, Cambridge. MR1820960

Mease, D., Wyner, A. and Buja, A. (2007). Boosted classification trees and class probability/quantile estimation. *J. Machine Learning Research* **8** 409–439.

Mease, D. and Wyner, A. (2007). Evidence contrary to the statistical view of boosting. *J. Machine Learning Research*. To appear.

Ridgeway, G. (1999). The state of boosting. *Comput. Sci. Statistics* **31** 172–181.

Ridgeway, G. (2000). Discussion of "Additive logistic regression: A statistical view of boosting," by J. Friedman, T. Hastie and R. Tibshirani. *Ann. Statist.* **28** 393–400. MR1790002

Schapire, R. E. and Singer, Y. (1999). Improved boosting algorithms using confidence-rated predictions. *Machine Learning* **37** 297–226. MR1811573

Wyner, A. (2003). On boosting and the exponential loss. In *Proceedings of the Ninth International Workshop on Artificial Intellingence and Statistics*.